\pdfoutput=1

\documentclass[11pt]{article}

\usepackage[preprint]{acl}

\usepackage{times}
\usepackage{latexsym}

\usepackage[T1]{fontenc}

\usepackage[utf8]{inputenc}

\usepackage{microtype}
\usepackage{booktabs}

\usepackage{inconsolata}

\usepackage{graphicx}
\usepackage{float}
\usepackage{overpic}
\usepackage{wrapfig}
\usepackage{caption}
\usepackage{subfig}
\usepackage{multirow}

\usepackage{amsmath}
\usepackage{amsthm}
\usepackage{amssymb}

\usepackage{bbding}
\usepackage{enumitem}

\title{MegaTTS 3: Sparse Alignment Enhanced Latent Diffusion Transformer \\ for Zero-Shot Speech Synthesis}

\author{
 \textbf{Ziyue Jiang\textsuperscript{1,2}\thanks{
Intern at ByteDance.}},
 \textbf{Yi Ren\textsuperscript{2}},
 \textbf{Ruiqi Li\textsuperscript{1,2}},
 \textbf{Shengpeng Ji\textsuperscript{1}},
 \textbf{Boyang Zhang\textsuperscript{1}},
 \textbf{Zhenhui Ye\textsuperscript{1}},
 \textbf{Chen Zhang\textsuperscript{2}},
\\
 \textbf{Bai Jionghao\textsuperscript{1}},
 \textbf{Xiaoda Yang\textsuperscript{1}},
 \textbf{Jialong Zuo\textsuperscript{1}},
 \textbf{Yu Zhang\textsuperscript{1}},
 \textbf{Rui Liu\textsuperscript{3}},
 \textbf{Xiang Yin\textsuperscript{2}},
 \textbf{Zhou Zhao\textsuperscript{1}}
\\
 \textsuperscript{1}Zhejiang University,
 \textsuperscript{2}ByteDance,
 \textsuperscript{3}Inner Mongolia University
\\
 \small{
   \href{ziyuejiang341@gmail.com}{ziyuejiang341@gmail.com},
   \href{zhaozhou@zju.edu.cn}{zhaozhou@zju.edu.cn}
 }
}

\begin{document}
\maketitle
\begin{abstract}
While recent zero-shot text-to-speech (TTS) models have significantly improved speech quality and expressiveness, 
mainstream systems still suffer from issues related to speech-text alignment modeling: 1) models without explicit speech-text alignment modeling exhibit less robustness, especially for hard sentences in practical applications; 2) predefined alignment-based models suffer from naturalness constraints of forced alignments. This paper introduces \textit{MegaTTS 3}, a TTS system featuring an innovative sparse alignment algorithm that guides the latent diffusion transformer (DiT). Specifically, we provide sparse alignment boundaries to MegaTTS 3 to reduce the difficulty of alignment without limiting the search space, thereby achieving high naturalness. Moreover, we employ a multi-condition classifier-free guidance strategy for accent intensity adjustment and adopt the piecewise rectified flow technique to accelerate the generation process. Experiments demonstrate that MegaTTS 3 achieves state-of-the-art zero-shot TTS speech quality and supports highly flexible control over accent intensity. Notably, our system can generate high-quality one-minute speech with only 8 sampling steps. Audio samples are available at~\url{https://sditdemo.github.io/sditdemo/}.
\end{abstract}


\section{Introduction}
In recent years, neural codec language models~\citep{wang2023neural,zhang2023speak,song2024ella,xin2024rall} and large-scale diffusion models ~\citep{shen2023naturalspeech,le2023Voicebox,lee2024ditto,eskimez2024e2,ju2024naturalspeech,yang2024simplespeech,yang2024simplespeech2} have brought considerable advancements to the field of speech synthesis. Unlike traditional text-to-speech (TTS) systems~\citep{shen2018natural,jia2018transfer,li2019neural,kim2020glow,ren2019fastspeech,kim2021conditional,kim2022guided}, these models are trained on large-scale, multi-domain speech corpora, which contributes to notable improvements in the naturalness and expressiveness of synthesized audio. Given only seconds of speech prompt, they can synthesize identity-preserving speech in a zero-shot manner.

To generate high-quality speech with clear and expressive pronunciation, a TTS model must establish an alignment mapping from text to speech signals~\citep{kim2020glow,tan2021survey}. However, from the perspective of speech-text alignment, current solutions suffer from the following issues:

\begin{itemize}
    \item \textbf{Models with implicit speech-text alignment} achieve the soft alignment paths through attention mechanisms~\citep{wang2023neural,chen2024vall,du2024cosyvoice}. These models can be categorized into: 1) autoregressive codec language models (AR LM), which are inefficient and lack robustness. The lengthy discrete speech codes, which typically require a bit rate of 1.5 kbps~\citep{kumar2024high,wu2024towards}, impose a significant burden on these autoregressive language models; 2) diffusion-based models without explicit duration modeling~\citep{lee2024ditto,eskimez2024e2,lovelace2023simple,gao2023e3,cambara2024mapache,yang2024simplespeech,yang2024simplespeech2}, which significantly speeds up the speech generation process. However, when compared with methods that adopt forced alignment, these models exhibit a decline in speech intelligibility. Besides, these methods cannot provide fine-grained control over the duration of specific pronunciations and can only adjust the overall speech rate.

    \item \textbf{Predefined alignment-based methods} have prosodic naturalness constraints of forced alignments. During training, alignment paths~\citep{ren2020fastspeech,kim2020glow} are directly introduced into their models~\citep{le2023Voicebox,shen2023naturalspeech,ju2024naturalspeech} to reduce the complexity of text-to-speech generation, which achieves higher intelligibility. Nevertheless, they suffer from the following limitations: 1) predefined alignments constrain the model's search space to produce more natural-sounding speech~\citep{anastassiou2024seed,chen2024vall}; 2) the overall naturalness is highly dependent on the performance of duration models.
\end{itemize}

Intuitively, we can integrate the two aforementioned diffusion-based methods to pursue optimal performance. To be specific, we propose a novel sparse speech-text alignment strategy to enhance the latent diffusion transformer (DiT), termed MegaTTS 3. In our approach, phoneme tokens are sparsely distributed within the corresponding forced alignment regions to provide coarse pronunciation information that is then refined by the latent DiT model. Experimental results demonstrate that MegaTTS 3 achieves nearly state-of-the-art speech intelligibility and speaker similarity on the LibriSpeech test-clean set~\citep{panayotov2015librispeech} with only 8 sampling steps, while also exhibiting high speech naturalness. The main contributions of this work are summarized as follows:

\begin{itemize}
\item We design a sparse alignment enhanced latent diffusion transformer model, which effectively integrates the strengths of the two aforementioned speech-text alignment approaches. Notably, our model also demonstrates greater robustness to duration prediction errors compared to methods with forced alignment. 

\item To achieve higher generation quality and more flexible control, we propose a multi-condition CFG strategy to adjust the guidance scales for speaker timbre and text content separately. Furthermore, we discover that the text guidance scale can also be used to modulate the intensity of personal accents, offering a new direction for enhancing speech expressiveness.

\item We successfully reduce the inference steps from 25 to 8 with the piecewise rectified flow (PeRFLow) technique, achieving highly efficient zero-shot TTS with minimal quality degradation. We also visualize the attention matrices across various layers of MegaTTS 3 and obtain insightful findings in Appendix~\ref{app:vis_diff_attn}.

\end{itemize}
\section{Background}

\paragraph{Zero-shot TTS.} 
Zero-shot TTS~\citep{casanova2022yourtts,wang2023neural,zhang2023speak,shen2023naturalspeech,le2023Voicebox,jiang2024mega,liu2024autoregressive,lee2024ditto,li2024styletts,lee2023hierspeech++,ju2024naturalspeech,meng2024autoregressive,chen2024f5} aims to synthesize unseen voices with speech prompts. Among them, neural codec language models~\citep{chen2024vall} are the first that can autoregressively synthesize speech that rivals human recordings in naturalness and expressiveness. However, they still face several challenges, such as the lossy compression in discrete audio tokenization and the time-consuming nature of autoregressive generation. To address these issues, some subsequent works explore solutions based on continuous vectors and non-autoregressive diffusion models~\citep{shen2023naturalspeech,le2023Voicebox,lee2024ditto,eskimez2024e2,yang2024simplespeech,yang2024simplespeech2,chen2024f5}. These works can be categorized into two main types: 1) the first type directly models speech-text alignments using attention mechanisms without explicit duration modeling~\citep{lee2024ditto,eskimez2024e2}. Although these models perform well in terms of generation speed and quality, their robustness, especially in challenging cases, still requires enhancement. The second category~\citep{shen2023naturalspeech,le2023Voicebox} utilizes predefined alignments to simplify alignment learning. However, the search space of the generated speech of these models is limited by predefined alignments. To address these limitations, we propose a sparse alignment mechanism to reduce the constraints of predefined alignment-based methods while also reducing the difficulty of speech-text alignment learning.

\begin{figure*}[!t]
	\centering
	\includegraphics[width=0.95\textwidth]{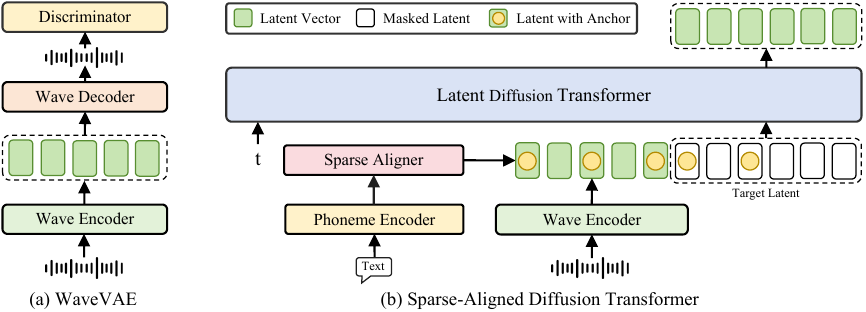}
	\caption{(a) The WaveVAE model; (b) Overview of our model. We insert the sparse alignment anchors into the latent vector sequence to provide coarse alignment information. The transformer blocks in MegaTTS 3 will automatically build fine-grained alignment paths.}
	\label{fig:arch_overview}
\end{figure*}

\paragraph{Accented TTS.} While accented TTS is not yet mainstream in the field of speech synthesis, it offers valuable potential for customized TTS services, by enhancing the expressiveness of speech synthesis systems and improving listeners' comprehension of speech content~\citep{tan2021survey,melechovsky2022accented,badlani2023multilingual,zhou2024multi,shah2024parrottts,ma2023accent,inoue2024macst,zhong2024accentbox}. With the emergence of conversational AI systems, accented TTS technology has even broader application scenarios. In this paper, we focus on a specific task of accented TTS: adjusting the accent intensity of speakers to make them sound like native English speakers or accented speakers who use English as a second language~\citep{liu2024controllable}. Unlike previous work, our approach does not require paired data or accurate accent labels; instead, it allows for flexible control over the accent intensity using the proposed multi-condition CFG mechanism. In addition, we describe the CFG mechanism used in zero-shot TTS systems in Appendix~\ref{app:CFG_in_zs_tts}.
\section{Method}
\label{method}
This section introduces MegaTTS 3. To begin with, we describe the architecture design of MegaTTS 3. Then, we provide detailed explanations of the sparse alignment mechanism, the piecewise rectified flow acceleration technique, and the multi-condition classifier-free guidance strategy.

\subsection{Architecture}
\label{method:main_arch}
\paragraph{WaveVAE.} As shown in Figure~\ref{fig:arch_overview} (a), given a speech waveform $s$, the VAE encoder $E$ encodes $s$ into a latent vector $z$, and the wave decoder $D$ reconstructs the waveform $x = D(z) = D(E(s))$. To reduce the computational burden of the model and simplify speech-text alignment learning, the encoder $E$ downsamples the waveform by a factor of $d$ in length. The encoder $E$ is similar to the one used in~\citet{ji2024wavtokenizer}, and the decoder $D$ is based on~\citet{kong2020hifi}. We also adopt the multi-period discriminator (MPD), multi-scale discriminator (MSD), and multi-resolution discriminator (MRD) ~\citep{kong2020hifi,jang2021univnet} to model the high-frequency details in waveforms, which ensure perceptually high-quality reconstructions. The training loss of the speech compression model can be formulated as $\mathcal{L} = \mathcal{L}_{\mathrm{rec}} + \mathcal{L}_{\mathrm{KL}} + \mathcal{L}_{\mathrm{Adv}}$, where $\mathcal{L}_{\mathrm{rec}}=\|s-\hat{s}\|^2$ is the spectrogram reconstruction loss, $\mathcal{L}_{\mathrm{KL}}$ is the slight KL-penalty loss~\citep{rombach2022high}, and $\mathcal{L}_{\mathrm{Adv}}$ is the LSGAN-styled adversarial loss~\citep{mao2017least}. After training, a one-second speech clip can be encoded into 25 vector frames. For more details, please refer to Appendix~\ref{app:model_config} and~\ref{app:evaluation_speech_compression}.

\paragraph{Latent Diffusion Transformer with Masked Speech Modeling.}
The latent diffusion transformer is used to predict speech that matches the style of the given speaker and the content of the provided text. Given the random variables $Z_{0}$ sampled from a standard Gaussian distribution $\pi_{0}$ and $Z_{1}$ sampled from the latent space given by the speech compression model with data density $\pi_{1}$, we adopt the rectified flow~\citet{liu2022flow} to implicitly learn the transport map $T$, which yields $Z_{1} := T(Z_{0})$. The rectified flow learns $T$ by constructing the following ordinary differential equation (ODE):
\begin{equation}
    \small
    \mathrm{d}Z_t = v(Z_t, t)\,\mathrm{d}t,
    \label{eq:1}
\end{equation}
where $t\in[0,1]$ denotes time and $v$ is the drift force. Equation~\ref{eq:1} converts $Z_{0}$ from $\pi_{0}$ to $Z_{1}$ from $\pi_{1}$. The drift force $v$ drives the flow to follow the direction $(Z_{1}-Z_{0})$. The latent diffusion transformer,  parameterized by $\theta$, can be trained by estimating $v(Z_{t}, t)$ with $v_{\theta}(Z_{t}, t)$ through minimizing the least squares loss with respect to the line directions $(Z_{1}-Z_{0})$:
\begin{equation}
    \small
    \min_v \int_0^1 \mathbb{E} \left[ \| (Z_1 - Z_0) - v(Z_t, t) \|^2 \right] \, \mathrm{d}t.
    \label{eq:2}
\end{equation}
We use the standard transformer block from LLAMA~\citep{dubey2024llama} as the basic structure for MegaTTS 3 and adopt the Rotary Position Embedding (RoPE)~\citep{su2024roformer} as the positional embedding. During training, we randomly divide the latent vector sequence into a prompt region $z_{prompt}$ and a masked target region $z_{target}$, with the proportion of $z_{prompt}$ being $\gamma \sim U(0.1, 0.9)$. We use $v_{\theta}$ to predict the masked target vector $\hat{z}_{target}$ conditioned on $z_{prompt}$ and the phoneme embedding $p$, denoted as $v_{\theta}(\hat{z}_{target}|z_{prompt}, p)$. The loss is calculated using only the masked region $z_{target}$. MegaTTS 3 learns the average pronunciation from $p$ and the specific characteristics such as timbre, accent, and prosody of the corresponding speaker from $z_{prompt}$.



\subsection{Sparse Alignment Enhanced Latent Diffusion Transformer (MegaTTS 3)}
\label{method:sec_3_2}
In this subsection, we describe the sparse alignment strategy as the foundation of MegaTTS 3, followed by the piecewise rectified flow and multi-condition CFG strategies to further enhance MegaTTS 3's capacity.

\paragraph{Sparse Alignment Strategy.}
Let’s first analyze the reasons behind the characteristics of different speech-text alignment modeling methods in depth. Implicitly modeling speech-text alignment is a relatively challenging task, which consequently leads to suboptimal speech intelligibility, particularly in hard cases. On the other hand, employing predefined hard alignment paths constrains the model's search space to produce more natural-sounding speech. The characteristics of these systems motivate us to design an approach that combines the advantages of both: we first provide a rough alignment to MegaTTS 3 and then use attention mechanisms in Transformer blocks to construct the fine-grained implicit alignment path. The visualizations of the implicit alignment paths are included in Appendix~\ref{app:vis_diff_attn}. In specific, denote the latent speech vector sequence as $z=[z_1, z_2, \cdots, z_n]$, the phoneme sequence as $p=[p_1, p_2, \cdots, p_m]$, and the phoneme duration sequence as $d=[d_1, d_2, \cdots, d_m]$, where $n$, $m$ is the length of the sequence. The length of the speech vector that corresponds to a phoneme $p_i$ is the duration $d_i$. Given $d=[2, 2, 3]$, the hard speech-text alignment path can be denoted as $a=[p_1, p_1, p_2, p_2, p_3, p_3, p_3]$. To construct the rough alignment $\tilde{a}$, we randomly retain only one anchor for each phoneme: $\tilde{a} = [\underline{M}, p_1, p_2, \underline{M}, \underline{M}, \underline{M}, P_3]$, where $\underline{M}$ represents the mask token. $\tilde{a}$ is downsampled with convolution layers to match the length of the latent sequence $z$. Then, we directly concatenate the downsampled $\tilde{a}$ and $z$ along the channel dimension. The anchors in $\tilde{a}$ provide MegaTTS 3 with approximate positional information for each phoneme, simplifying the learning process of speech-text alignment. At the same time, the rough alignment information does not limit MegaTTS 3's search space and also enables fine-grained control over each phoneme's duration.

\paragraph{Piecewise Rectified Flow Acceleration.}
We adopt Piecewise Rectified Flow (PeRFlow)~\citep{yan2024perflow} to distill the pretrained MegaTTS 3 model into a more efficient generator.
Although our MegaTTS 3 is non-autoregressive in terms of the time dimension, it requires multiple iterations to solve the Flow ODE. The number of iterations (i.e., number of function evaluations, NFE) has a great impact on inference efficiency, especially when the model scales up further. Therefore, we adopt the PeRFlow technique to further reduce NFE by segmenting the flow trajectories into multiple time windows. Applying reflow operations within these shortened time intervals, PeRFlow eliminates the need to simulate the full ODE trajectory for training data preparation, allowing it to be trained in real-time alongside large-scale real data during the training process. Given number of windows $K$, we divide the time $t\in[0,1]$ into $K$ time windows $\{ (t_{k-1}, t_{k}] \}^{K}_{k=1}$. Then, we randomly sample $k\in\{1,\cdots,K\}$ uniformly. We use the startpoint of the sampled time window $z_{t_{k-1}} = \sqrt{1 - \sigma^2(t_{k-1})} z_1 + \sigma(t_{k-1}) \epsilon$ to solve the endpoint of the time window $\hat{z}_{t_k} = \phi_{\theta}(z_{t_{k-1}}, t_{k-1}, t_{k})$, where $\epsilon \sim \mathcal{N}(0, I) $ is the random noise, $\sigma(t)$ is the noise schedule, and $\phi_{\theta}$ is the ODE solver of the teacher model. Since $z_{t_{k-1}}$ and $\hat{z}_{t_{k}}$ is available, the student model $\hat{\theta}$ can be trained via the following objectives:
\begin{equation}
    \small
    \ell = \left\lVert v_{\hat{\theta}}(z_t, t) - \frac{\hat{z}_{t_k} - z_{t_{k-1}}}{t_k - t_{k-1}} \right\lVert^2,
    \label{eq:3}
\end{equation}
where $v_{\hat{\theta}}$ is the estimated drift force with parameter $\hat{\theta}$ and $t$ is uniformly sampled from $(t_{k-1},t_{k}]$. We provide details of PeRFlow training for MegaTTS 3 in Appendix~\ref{app:details_perflow_training}.

\paragraph{Multi-condition Classifier-Free Guidance (CFG).}
We employ classifier-free guidance approach~\citep{ho2022classifier} to steer the model $g_{\theta}$'s output towards the conditional generation $g_{\theta}(z_t,c)$ and away from the unconditional generation $g_{\theta}(z_t,\varnothing)$:
\begin{equation}
    \small
    \hat{g}_{\theta}(z_t, c) = g_{\theta}(z_t, \varnothing) + \alpha \cdot \left[ g_{\theta}(z_t, c) - g_{\theta}(z_t, \varnothing) \right],
    \label{eq:4}
\end{equation}
where $c$ denotes the conditional state, $\varnothing$ denotes the unconditional state, and $\alpha$ is the guidance scale selected based on experimental results. Unlike standard classifier-free guidance, MegaTTS 3's conditional states $c$ consist of two components: phoneme embeddings $p$ and the speaker prompt $z_{prompt}$. In the experiments, as the text guidance scale increases, we observe that the pronunciation changes according to the following pattern: 1) starting with improper pronunciation; 2) then shifting to pronouncing with the current speaker's accent; 3) and finally approaching the standard pronunciation of the target language. The detailed experimental setup is described in Appendix~\ref{app:additional_detials_for_mt_cfg}. This allows us to use the text guidance scale $\alpha_{txt}$ to control the accent intensity. At the same time, the speaker guidance scale $\alpha_{spk}$ should be a relatively high value to ensure a high speaker similarity. Therefore, we adopt the multi-condition classifier-free guidance technique to separately control $\alpha_{txt}$ and $\alpha_{spk}$:
\begin{equation}
    \small
    \begin{split}
    \hat{g}_{\theta}(z_t, p, z_{prompt}) = & \alpha_{spk} \left[ g_{\theta}(z_t, p, z_{prompt}) - g_{\theta}(z_t, p, \varnothing) \right] \\
    & + \alpha_{txt} \left[ g_{\theta}(z_t, p, \varnothing) - g_{\theta}(z_t, \varnothing, \varnothing) \right] \\ 
    & + g_{\theta}(z_t, \varnothing, \varnothing) \\
    \end{split}
    \label{eq:5}
\end{equation}
In training, we randomly drop condition $z_{prompt}$ with a probability of $p_{spk} = 0.10$. Only when $z_{prompt}$ is dropped, we randomly drop condition $p$ with a probability of 50\%. Therefore, our model is able to handle all three types of conditional inputs described in Equation~\ref{eq:5}. We select the guidance scale $\alpha_{spk}$ and $\alpha_{txt}$ based on experimental results.

\begin{table*}[!t]
\small
\centering
\begin{tabular}{@{}l|cc|ccc|cc|c@{}}
\toprule
\bfseries Model & \bfseries \#Params & \bfseries Training Data & \bfseries SIM-O$\uparrow$ & \bfseries SIM-R$\uparrow$ & \bfseries WER$\downarrow$ & \bfseries CMOS$\uparrow$& \bfseries SMOS$\uparrow$ & \bfseries RTF$\downarrow$ \\      
\midrule
GT & - & - & 0.68 & - & 1.94\% & +0.12 & 3.92 & - \\
\midrule
VALL-E 2$^{*}$ & 0.4B & LibriHeavy & 0.64 & 0.68 & 2.44\% & - & - & - \\
VoiceBox$^{\dag}$ & 0.4B & Collected (60kh) &  0.64 & 0.67 & 2.03\% & -0.20 & 3.81 & 0.340 \\
DiTTo-TTS$^{*}$ & 0.7B & Collected (55kh) & 0.62 & 0.65 & 2.56\% & - & - & - \\
NaturalSpeech 3$^{\dag}$ & 0.5B & LibriLight & 0.67 & 0.76 & \bfseries 1.81\% & -0.10 & 3.95 & 0.296\\
CosyVoice & 0.4B & Collected (172kh) & 0.62 & - & 2.24\% & -0.18 & 3.93 & 1.375\\
MaskGCT & 1.0B & Emilia (100kh) & 0.69 & - & 2.63\% & - & - & - \\
F5-TTS & 0.3B & Emilia (100kh) & 0.66 & - & 1.96\% & -0.12 & 3.96 & 0.307 \\
\midrule
MegaTTS 3 & 0.3B & LibriLight & \bfseries 0.71 & \bfseries 0.78 & \bfseries 1.82\%  & \bfseries 0.00 & \bfseries 3.98 & 0.188 \\
MegaTTS 3-accelerated & 0.3B & LibriLight & 0.70 & \bfseries0.78 & 1.86\%  & -0.03  & 3.96 & \bfseries 0.124 \\

\bottomrule
\end{tabular}
\caption{Zero-shot TTS results on the LibriSpeech test-clean set following NaturalSpeech 3~\citep{ju2024naturalspeech}. $^{*}$ means the results are obtained from the paper. $^{\dag}$ means the
results are obtained from the authors. \#Params denotes the number of parameters. RTF denotes the real-time factor.}
\label{table:en_zs_tts}
\end{table*}

\begin{table}[!t]
\small
\centering
\begin{tabular}{@{}l|ccc@{}}
\toprule
\bfseries Model & \bfseries \#Params & \bfseries SIM-O$\uparrow$ & \bfseries WER$\downarrow$  \\       
\midrule
GT              & - &  0.69 & 2.23\%   \\
\midrule
CosyVoice       & 0.3B & 0.66 & 3.59\% \\
E2 TTS          & 0.3B & 0.69 & 2.95\% \\
F5-TTS          & 0.3B & 0.66 & 2.42\% \\
\midrule
MegaTTS 3           & 0.3B & \bfseries 0.70 & \bfseries 2.31\% \\
\bottomrule
\end{tabular}
\caption{Zero-shot TTS results on the LibriSpeech-PC test-clean set following F5-TTS~\citep{chen2024f5}. \#Params denotes the number of parameters.}
\label{table:en_zs_tts_valle}
\end{table}

\section{Experiments} 
In this subsection, we describe the datasets, training, inference, and evaluation metrics. We provide the model configuration and detailed hyper-parameter setting in Appendix~\ref{app:model_config}.

\subsection{Experimental setup}
\label{Experimental_Setup}
\paragraph{Datasets.} We train MegaTTS 3 on the LibriLight~\citep{kahn2020libri} dataset, which contains 60k hours of unlabeled speech derived from LibriVox audiobooks. All speech data are sampled at 16KHz. We transcribe the speeches using an internal ASR system and extract the predefined speech-text alignment using the external alignment tool~\citep{mcauliffe2017montreal}. We utilize three benchmark datasets: 1) the librispeech~\citep{panayotov2015librispeech} test-clean set following NaturalSpeech 3~\citep{ju2024naturalspeech} for zero-shot TTS evaluation; 2) the LibriSpeech-PC test-clean set following F5-TTS~\citep{chen2024f5} for zero-shot TTS evaluation; 3) the L2-arctic dataset~\citep{zhao2018l2arctic} following~\citep{melechovsky2022accented,liu2024controllable} for accented TTS evaluation.

\paragraph{Training and Inference.} 
We train the WaveVAE model and MegaTTS 3 on 8 NVIDIA A100 GPUs. The batch sizes, optimizer settings, and learning rate schedules are described in Appendix~\ref{app:model_config}. It takes 2M steps for the WaveVAE model's training and 1M steps for MegaTTS 3's training until convergence. The pre-training of MegaTTS 3 requires 800k steps and PeRFlow distillation requires 200k steps.
\paragraph{Objective Metrics.} 1) For zero-shot TTS, we evaluate speech intelligibility using the word error rate (WER) and speaker similarity using SIM-O~\citep{ju2024naturalspeech}. To measure SIM-O, we utilize the WavLM-TDCNN speaker embedding model\footnote{\url{https://github.com/microsoft/UniSpeech/tree/main/downstreams/speaker_verification}} to calculate the cosine similarity score between the generated samples and the prompt. As SIM-R~\citep{le2023Voicebox} is not comparable across baselines using different acoustic tokenizers, we recommend focusing on SIM-O in our experiments. The similarity score is in the range of $\left[-1,1\right]$, where a higher value indicates greater similarity. In terms of WER, we use the publicly available HuBERT-Large model~\citep{hsu2021hubert}, fine-tuned on the 960-hour LibriSpeech training set, to transcribe the generated speech. The WER is calculated by comparing the transcribed text to the original target text. All samples from the test set are used for the objective evaluation; 2) For accented TTS, we evaluate the Mel Cepstral Distortion (MCD) in dB level and the moments (standard deviation ($\sigma$), skewness ($\gamma$) and kurtosis ($\kappa$))~\citep{andreeva2014differences,niebuhr2019measuring} of the pitch distribution to evaluate whether the model accurately captures accent variance. 

\paragraph{Subjective Metrics.} We conduct the MOS (mean opinion score) evaluation on the test set to measure the audio naturalness via Amazon Mechanical Turk. We keep the text content and prompt speech consistent among different models to exclude other interference factors. We randomly choose 40 samples from the test set of each dataset for the subjective evaluation, and each audio is listened to by at least 10 testers. We analyze the MOS in three aspects: CMOS (quality, clarity, naturalness, and high-frequency details), SMOS (speaker similarity in terms of timbre reconstruction and prosodic pattern), and ASMOS (accent similarity). We tell the testers to focus on one corresponding aspect and ignore the other aspect when scoring.

\subsection{Results of Zero-Shot Speech Synthesis}
\label{exp_zero-shot}

\paragraph{Evaluation Baselines.} We compare the zero-shot speech synthesis performance of MegaTTS 3 with 11 strong baselines, including: 1) VALL-E 2~\citep{chen2024vall}; 2) VoiceBox~\citep{le2023Voicebox}; 3) DiTTo-TTS~\citep{lee2024ditto}; 4) NaturalSpeech 3~\citep{ju2024naturalspeech}; 5) CosyVoice~\citep{du2024cosyvoice}; 6) MaskGCT~\citep{wang2024maskgct}; 7) F5-TTS~\citep{chen2024f5}; 8) E2 TTS~\citep{eskimez2024e2}. Explanation and details of the selected baseline systems are provided in Appendix~\ref{app:detail_zs_tts_baseline}.

\paragraph{Analysis} 
As shown in Table~\ref{table:en_zs_tts}, we can see that 1) MegaTTS 3 achieves state-of-the-art SIM-O, SMOS, and WER scores, comparable to NaturalSpeech 3 (the counterpart with forced alignment), and significantly surpasses other baselines without explicit alignments. The improved SIM-O and SMOS suggest that the proposed sparse alignment effectively simplifies the text-to-speech mapping challenge like predefined forced duration information, allowing the model to focus more on learning timbre information. And the improved WER indicates that MegaTTS 3 also enjoys strong robustness; 2) MegaTTS 3 significantly surpasses all baselines in terms of CMOS, demonstrating the effectiveness of the proposed sparse alignment strategy; 3) After the PeRFlow acceleration, the student model of MegaTTS 3 shows on par quality with the teacher model and enjoys fast inference speed. We also conduct the experiments on the LibriSpeech-PC test-clean set provided by F5-TTS and the results are shown in Table~\ref{table:en_zs_tts_valle}, which also demonstrates that our method achieves state-of-the-art performance in terms of speaker similarity and speech intelligibility. The duration controllability of MegaTTS 3 is verified in Appendix~\ref{app:dur_contol}. In the demo page, we also demonstrate that our method can maintain high naturalness even when the performance of the duration predictor is suboptimal (while MegaTTS 3 with forced alignment fails).

\begin{table*}[!t]
\small
\centering
\begin{tabular}{@{}l|cccc|ccc@{}}
\toprule
\bfseries Model & \bfseries MCD (dB) $\downarrow$ & \bfseries $\sigma$ $\uparrow$ & \bfseries $\gamma$ $\downarrow$ & \bfseries $\kappa$ $\downarrow$ & \bfseries ASMOS $\uparrow$ & \bfseries CMOS $\uparrow$ & \bfseries SMOS $\uparrow$ \\       
\midrule
GT      &   - & 45.1 & 0.591 & 0.783 & 4.03 & +0.09 & 3.95 \\

CTA-TTS & 5.98  & 41.1 & 0.602 & 0.799 & 3.72 & -0.60 & 3.64 \\

MegaTTS 3   & \bfseries 5.69 & \bfseries 42.3  & \bfseries 0.601  & \bfseries 0.790  & \bfseries 3.84 & \bfseries +0.00 & \bfseries 3.89 \\

\bottomrule
\end{tabular}
\caption{The objective and subjective experimental results for accented TTS. MCD (dB) denotes the Mel Cepstral Distortion at the dB level. $\sigma$, $\gamma$, and $\kappa$ are the standard deviation, skewness, and kurtosis of the pitch distribution.}
\label{table:accent-tts-result}
\end{table*}

\begin{table}[!t]
\small
\centering
\begin{tabular}{@{}l|cccccc@{}}
\toprule
\bfseries Method & \bfseries MCD$\downarrow$ & \bfseries GPE$\downarrow$ & \bfseries VDE$\downarrow$ & \bfseries FFE$\downarrow$ \\       
\midrule
NaturalSpeech 3      &  4.45  & 0.44 & 0.33 & 0.37\\
Ours w/ F.A. &  4.48   & 0.44 & 0.35 & 0.40 \\
\midrule
Ours w/ S.A. & \bfseries 4.42 & \bfseries 0.31 & \bfseries 0.29 & \bfseries 0.34 \\

\bottomrule
\end{tabular}
\caption{Comparisons about prosodic naturalness metrics on LibriSpeech test-clean set. ``F.A.'' denotes forced alignment and ``S.A.'' denotes sparse alignment.}
\label{table:expressiveness_exp_40}
\end{table}

\subsection{Experiments of Prosodic Naturalness}
\label{exp:prosodic_naturalness}
We also measure the objective metrics MCD, SSIM, STOI, GPE, VDE, and FFE following InstructTTS~\citep{yang2024instructtts} to evaluate the prosodic naturalness of our method. The results are presented in Table~\ref{table:expressiveness_exp_40}. Specifically, our method with sparse alignment (Ours w/ S.A.) achieves the best performance across all metrics, with an MCD of 4.42, GPE of 0.31, VDE of 0.29, and FFE of 0.34. These results indicate a significant improvement in prosodic naturalness compared to the baseline NaturalSpeech 3 and our method with forced alignment (Ours w/ F.A.), further validating the effectiveness of our sparse alignment strategy. Our method provides a noval and effective solution for speech synthesis applications that require high robustness and exceptional expressiveness, such as audiobook narration and virtual assistants.

\subsection{Results of Accented TTS}
\begin{figure}[!t]
  \centering
    \includegraphics[scale=0.50]{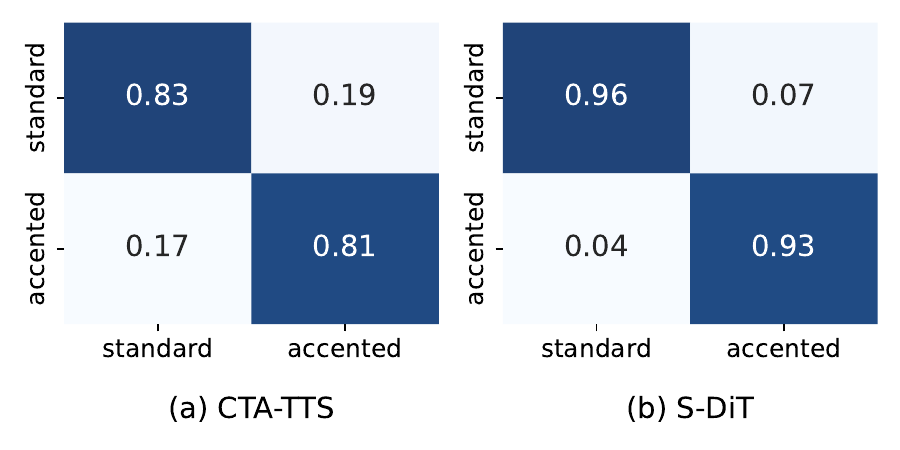}
  \caption{The confusion matrices between the perceived and intended accent categories of synthesized speech. The X-axis and Y-axis represent the intended and perceived categories, respectively.}
  \label{exp:accent_confusion_matrices}
\end{figure}

In this subsection, we evaluate the accented TTS performance of our model on the L2-ARCTIC dataset~\citep{zhao2018l2arctic}. This corpus includes recordings from non-native speakers of English whose first languages are Hindi, Korean, etc. In this experiment, we focus on verifying whether our model and baseline can synthesize natural speech with different accent types (standard English or English with specific accents) while maintaining consistent vocal timbre. We compare our MegaTTS 3 model with CTA-TTS~\citep{liu2024controllable}. More details of the baseline model are provided in Appendix~\ref{app:detail_accented_tts_experiment}. 1) First, we evaluate whether the models can synthesize high-quality speeches with accents. As shown in Table~\ref{table:accent-tts-result}, our MegaTTS 3 model significantly outperforms the CTA-TTS baseline in terms of the subjective accent similarity MOS core, the MCD (dB) values, and the statistical moments ($\sigma$, $\gamma$, and $\kappa$) of pitch distributions. These results demonstrate the superior accent learning capability of MegaTTS 3 compared to the baseline system. Besides, the MegaTTS 3 model achieves higher CMOS and SMOS scores compared to CTA-TTS, indicating a significant improvement in speech quality and speaker similarity; 2) Secondly, we evaluate whether the models can accurately control the accent types of the generated speeches. We follow CTA-TTS to conduct the intensity classification experiment~\citep{liu2024controllable}. At
run-time, we generate speeches with two accent types, and the listeners are instructed to classify the perceived accent categories, including ``standard'' and ``accented''. Figure~\ref{exp:accent_confusion_matrices} shows that our MegaTTS 3 significantly surpasses CTA-TTS in terms of accent controllability.

\begin{table*}[!ht]
\centering
\small
\begin{tabular}{@{}l|ccccccccc@{}}
\toprule
\textbf{Models} & \textbf{Tokens/s} & \textbf{Latent Layer} & \textbf{Type} & \textbf{PESQ$\uparrow$} & \textbf{STOI$\uparrow$} & \textbf{ViSQOL$\uparrow$} & \textbf{MCD$\downarrow$} & \textbf{UTMOS$\uparrow$} \\
\midrule
Encodec &  600 & 8 & Discrete & 3.16 & 0.94 & 4.31 & 1.63 & 3.07 \\
DAC &  450 & 9 & Discrete & \textbf{4.13} & \textbf{0.97} & \underline{4.68} & \underline{1.05} & 4.01 \\
WavTokenizer &  75 & 1 & Discrete & 2.55 & 0.88 & 3.83 & 1.99 & 4.07 \\
X-codec2 &  50 & 1 & Discrete & 3.03 & 0.91 & 4.12 & 1.72 & \textbf{4.13} \\
\midrule
WaveVAE & 25 & 1 & Continuous & \underline{3.84} & \underline{0.96} & \textbf{4.71} & \textbf{1.03} & \underline{4.10} \\
\bottomrule
\end{tabular}
\caption{Comparison of the reconstruction quality. The sampling rate are set to 16 kHz.  \textbf{Bold} and \underline{Underline} values indicate the best and second best results. ``Tokens/s'' means how many tokens a one-second speech will be compressed into.}
\label{app:table_recon_speech_compression}
\end{table*}

\begin{table}[!ht]
\small
\centering
\begin{tabular}{@{}l|cc@{}}
\toprule
\bfseries Setting & \bfseries SIM-O$\uparrow$ & \bfseries WER$\downarrow$ \\       
\midrule
Ours   & \bfseries 0.71           & \bfseries 1.82\%                 \\
\midrule
\textit{w/ Encodec}  & 0.56           & 2.24\%                 \\
\textit{w/ DAC}  &  0.64 & 1.93\%        \\
\bottomrule
\end{tabular}
\caption{Comparison of zero-shot TTS performance of MegaTTS 3 using different speech compression models on the LibriSpeech test-clean set.}
\label{app:table_zs_tts_different_codec}
\end{table}

\subsection{Evaluation of WaveVAE}
\label{app:evaluation_speech_compression}
First, we evaluate the reconstruction quality of the WaveVAE model, with results presented in Table~\ref{app:table_recon_speech_compression}. We report the objective metrics, including Perceptual Evaluation of Speech Quality (PESQ), Virtual Speech Quality Objective Listener (ViSQOL), and Mel-Cepstral Distortion (MCD). We select the following codec models as baselines: 1) EnCodec~\citep{defossez2022high}, a representative and pioneering work in the field of speech codec; 2) DAC~\citep{kumar2024high}, a high-bitrate audio codec model with high reconstruction quality; 3) WavTokenizer~\citep{ji2024wavtokenizer}, a low-bitrate speech codec model that focuses more on perceptual reconstruction quality; 4) X-codec2~\citep{ye2025llasa}, a low-bitrate speech codec model, leveraging the representations of a pre-trained model to further enhance overall quality. The results demonstrates that, despite applying higher compression rate, our solution achieves superior performance on various reconstruction metrics, such as MCD and ViSQOL. 

Second, to demonstrate the impact of different speech compression models on the overall performance of the TTS system, we extracted the latents from Encodec and DAC, respectively, for training our MegaTTS 3 model. We report the experimental results in Table~\ref{app:table_zs_tts_different_codec}. It can be seen that our method outperforms ``w/ DAC'' and ``w/ Encodec'', due to the fact that the latent space of our speech compression model is more compact (only 25 tokens per second). The results demonstrate the importance of our WaveVAE, a high-compression, high-reconstruction-quality speech codec model, for TTS systems. This conclusion is also verified by a previous work~\citep{lee2024ditto}, which shows compact target latents facilitate learning in diffusion models.

\subsection{Ablation Studies}
\label{exp:ablation_studies}
\begin{table}[!t]
\small
\centering
\begin{tabular}{@{}l|cc|cc@{}}
\toprule
\bfseries Setting & \bfseries SIM-O$\uparrow$ & \bfseries WER$\downarrow$ & \bfseries CMOS$\uparrow$ & \bfseries SMOS$\uparrow$ \\       
\midrule
Ours & 0.71 & 1.82$\%$ & 0.00 & 3.94 \\
\midrule
\textit{w/ F.A.}  & 0.70 & 1.80\% & -0.17 & 3.94 \\
\textit{w/o A.}        & 0.67 & 2.14\% & -0.12 & 3.88 \\
\midrule
\textit{w/ CFG}       & 0.68 & 1.79\% & -0.02 & 3.89 \\
\textit{w/o CFG}              & 0.43 & 6.85\% & -0.56 & 3.35  \\
\bottomrule
\end{tabular}
\caption{Ablation studies of alignment strategies and CFG mechanisms on the LibriSpeech test-clean set.}
\label{table:ablation_alignments_cfg}
\end{table}

We test the following four settings: 1) \textit{w/ FA}, which replaces the sparse alignment in MegaTTS 3 with forced alignment used in ~\citep{le2023Voicebox,shen2023naturalspeech}; 2) \textit{w/o A.}, we do not use the predefined alignments and modeling the duration information implicitly; 3) \textit{w/ CFG}, we use the standard CFG following the common practice in Diffusion-based TTS; 4) \textit{w/o CFG}, we do not use the CFG mechanism. All tests follow the experimental setup described in Section~\ref{exp_zero-shot}. The results are shown in Table~\ref{table:ablation_alignments_cfg}. For settings 1) and 2), it can be observed that both forced alignment and sparse alignment can enhance the performance of speech synthesis models. However, compared to forced alignment, sparse alignment does not constrain the model's search space, leading to a prosodic naturalness (see Section~\ref{exp:prosodic_naturalness}). Therefore, the sparse alignment strategy achieves $+0.17$ CMOS compared to the forced alignment strategy. For setting 3), compared with the standard CFG, our multi-condition CFG performs slightly better as it allows for flexible control over the weights between the text prompt and the speaker prompt. Setting 4) proves that the CFG mechanism is crucial for MegaTTS 3. Additionally, we visualize the attention score matrices from different transformer layers in MegaTTS 3 in Appendix~\ref{app:vis_diff_attn}, leading to some interesting observations.

\section{Conclusions}
In this paper, we introduce MegaTTS 3, a zero-shot TTS framework that leverages novel sparse alignment boundaries to ease the difficulty of alignment learning while retaining the naturalness of the generated speeches. This strategy allows MegaTTS 3 to combine the strengths of methods with both implicit alignments and predefined hard alignments. Additionally, we employ the PeRFlow technique to further accelerate the generation process and design a multi-condition CFG strategy to offer more flexible control over accents. Experimental results show that MegaTTS 3 achieves state-of-the-art zero-shot TTS speech quality while maintaining a more efficient pipeline. Moreover, the sparse alignment strategy also shows enhanced prosodic naturalness and higher robustness against a suboptimal duration predictor. Due to space constraints, further discussions are provided in the appendix.

\section*{Limitations}
\label{app:limitation_and_future_work}
In this section, we discuss the limitations of the proposed method and outline potential strategies for addressing them in future research. 
\begin{itemize}
\item \textbf{Language Coverage.} Although our model currently supports both English and Chinese, there are far more languages in the world. We plan to incorporate additional training data from a wider range of languages and apply adaptation-based techniques, such as LoRA tuning~\citep{hu2021lora}, to enhance speech quality for low-resource languages.
\item \textbf{Function Coverage.} We can make MegaTTS 3 more user-friendly by enabling it to generate speech in various styles according to text descriptions through instruction-based fine-tuning. We can further fine-tune MegaTTS 3 on the paralinguistic corpus, allowing it to generate speech that is closer to a natural human style.
\end{itemize}

\bibliography{custom}

\appendix

\section{Detailed Experimental Settings}

\subsection{Model Configuration}
\label{app:model_config}

\begin{itemize}
\item \textbf{The WaveVAE model} consists of a VAE encoder, a wave decoder, and discriminators; The VAE encoder follows the architecture used in~\citep{ji2024wavtokenizer}. The wave decoder consists of a convolutional upsampler and a Hifi-GAN decoder~\citep{kong2020hifi}. The latent channel size is set to 32. The weight of the KL loss is set to $1\times10^{-3}$, which only imposes a slight KL penalty on the learned latent. In training, we use batches of fixed length, consisting of 72,000 wavform frames, with a batch size set to 40 for each GPU. We use the Adam optimizer with a learning rate of $1\times10^{-4}$, $\beta_{1} = 0.9$, $\beta_{2} = 0.999$, and 10K warmup steps.

\item \textbf{The MegaTTS 3 model} use the standard transformer block from LLAMA~\citep{dubey2024llama} as the basic structure, which comprises a 24-layer Transformer with 16 attention heads and 1024 embedding dimensions. It contains 339M parameters in total. We adopt the Rotary Position Embedding (RoPE)~\citep{su2024roformer} as the positional embedding following the common practice in LLAMA implementations. For simplicity, we do not use the phoneme encoder and style encoder like previous works. We only use a linear projection layer to transform these features to the same dimension. During training, we use 8 A100 80GB GPUs with a batch size of 10K latent frames per GPU for 1M steps. We use the Adam optimizer with a learning rate of $5\times10^{-5}$, $\beta_{1} = 0.9$, $\beta_{2} = 0.999$, and 10K warmup steps. In zero-shot TTS experiments, we set the text guidance scale $\alpha_{txt}$ and the speaker guidance scale $\alpha_{spk}$ to 2.5 and 3.5, respectively. In accented TTS experiments, we set $\alpha_{spk}=6.5$, $\alpha_{txt}=1.5$ to generate the accented speech and set $\alpha_{spk}=2.0$, $\alpha_{txt}=5.0$ to generate the speech with standard English.

\end{itemize}

\subsection{Random Seeds}
\label{app:random_seeds}
We ran objective experiments 10 times with 10 different random seeds and obtained the averaged results. The chosen random seeds are [4475, 5949, 6828, 6744, 3954, 3962, 6837, 1237, 3824, 3163].

\subsection{Sampling Strategy}
For MegaTTS 3, we applied the Euler sampler with a fixed step size following the common practice in flow ODE sampling. We use 25 and 8 sampling steps for \textit{MegaTTS 3} and \textit{MegaTTS 3-accelerated}, respectively.

\subsection{Details about Zero-Shot TTS Baselines}
\label{app:detail_zs_tts_baseline}
In this subsection, we provide the details about the baselines in our zero-shot TTS experiments:

\begin{itemize}

\item \textbf{VALL-E 2}~\citep{chen2024vall}, based on VALL-E, introduces Repetition Aware Sampling to stabilize the decoding process and proposes the Grouped Code Modeling to effectively address the challenges of long sequence modeling.

\item \textbf{VoiceBox}~\citep{le2023Voicebox} is a non-autoregressive flow-matching model designed to infill mel-spectrograms based on provided speech context and text. We obtained the samples by contacting the authors.

\item \textbf{DiTTo-TTS}~\citep{lee2024ditto} addresses
the challenge of text-speech alignment via cross-attention mechanisms with the prediction of the total length of speech representations. We directly obtain the results of objective evaluations from their paper.

\item \textbf{NaturalSpeech 3}~\citep{ju2024naturalspeech} designs a neural codec with factorized vector quantization (FVQ) to disentangle speech waveform into subspaces of content, prosody, timbre, and acoustic details and propose a factorized diffusion model to generate attributes in each subspace following its corresponding prompt. We obtained the samples by contacting the authors.

\item \textbf{CosyVoice}~\citep{du2024cosyvoice} utilizes an LLM for text-to-token generation and a conditional flow matching model for token-to-speech synthesis. We use the official code and the model snapshot named ``CosyVoice-300M'' in our experiments\footnote{\url{https://github.com/FunAudioLLM/CosyVoice}}.

\item \textbf{MaskGCT}~\citep{wang2024maskgct} proposes a fully non-autoregressive codec-based TTS model that eliminates the need for explicit alignment information between text
and speech supervision, as well as phone-level duration prediction. We directly obtain the results of objective evaluations from their paper.

\item \textbf{F5-TTS}~\citep{chen2024f5} proposes a fully non-autoregressive text-to-speech system
based on flow matching with Diffusion Transformer (DiT). We use the official code and pretrained model in our experiments\footnote{\url{https://github.com/SWivid/F5-TTS}}.

\item \textbf{E2 TTS}~\citep{eskimez2024e2} proposes an easy non-autoregressive zero-shot TTS system, that offers human-level naturalness and state-of-the-art speaker similarity and intelligibility. We use the code implemented by F5-TTS authors in our experiments\footnote{\url{https://github.com/SWivid/F5-TTS}}.
\end{itemize}

The evaluation is conducted on a server with 1 NVIDIA V100 GPU and batch size 1. RTF denotes the real-time factor, i.e., the seconds required for the system (together with the vocoder) to synthesize one-second audio.

\subsection{Details about the Accented TTS Baseline}
\label{app:detail_accented_tts_experiment}
CTA-TTS~\citep{liu2024controllable} is a TTS framework that uses a phoneme recognition model to quantify the accent intensity in phoneme level for accent intensity control. CTA-TTS first trains the phoneme recognition model on the standard pronunciation LibriSpeech dataset, and then uses the output probability distribution of the model to assess the accent intensity and create accent labels on the accented L2Arctic dataset. These labels were input into the TTS model to enable control over accent intensity.

Systems like CTA-TTS require precise accent annotations during training, so we trained them on the L2-ARCTIC dataset. However, our model does not require accent annotations and learns different accent patterns from large-scale data, using only the multi-condition CFG mechanism to achieve accent intensity control. Therefore, we directly compare the zero-shot results of our model with the baselines, which is a more challenging task.

\subsection{Details in Subjective Evaluations}
\label{details_subjective_evaluation}
\begin{figure*}[!ht]
    \centering
	\begin{minipage}{0.85\linewidth}
		\centering
		\includegraphics[width=1\linewidth]{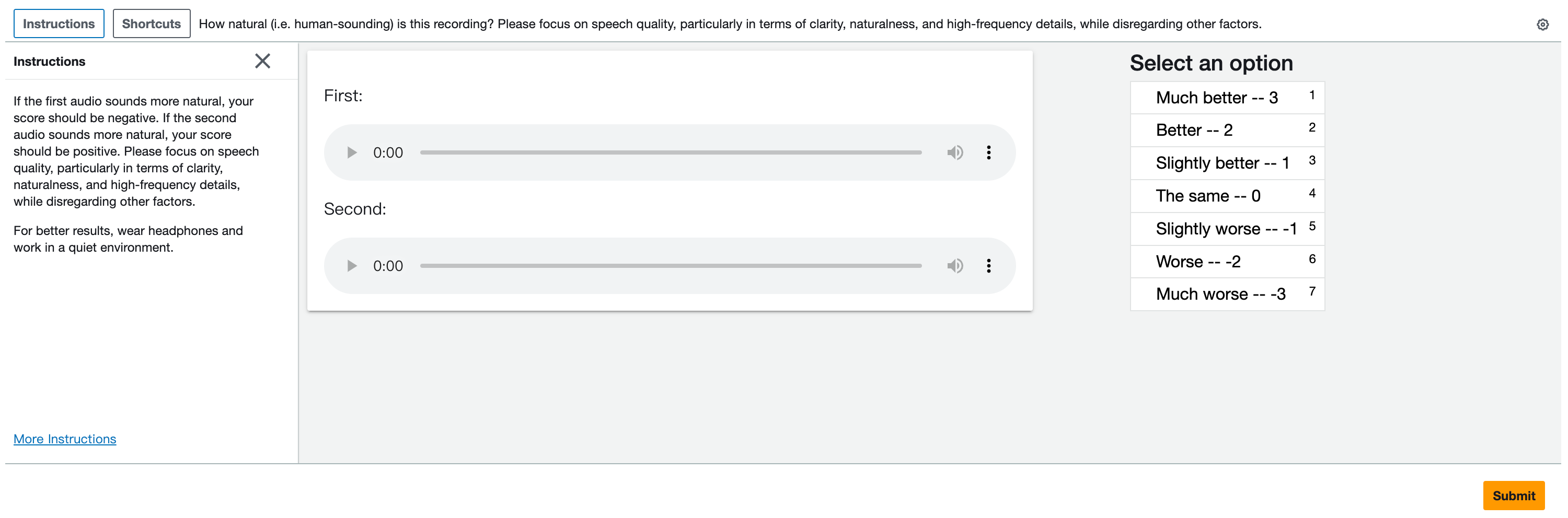}
		\caption*{(a)  Screenshot of CMOS testing.}
	\end{minipage}
	\centering
	\begin{minipage}{0.85\linewidth}
		\centering
		\includegraphics[width=1\linewidth]{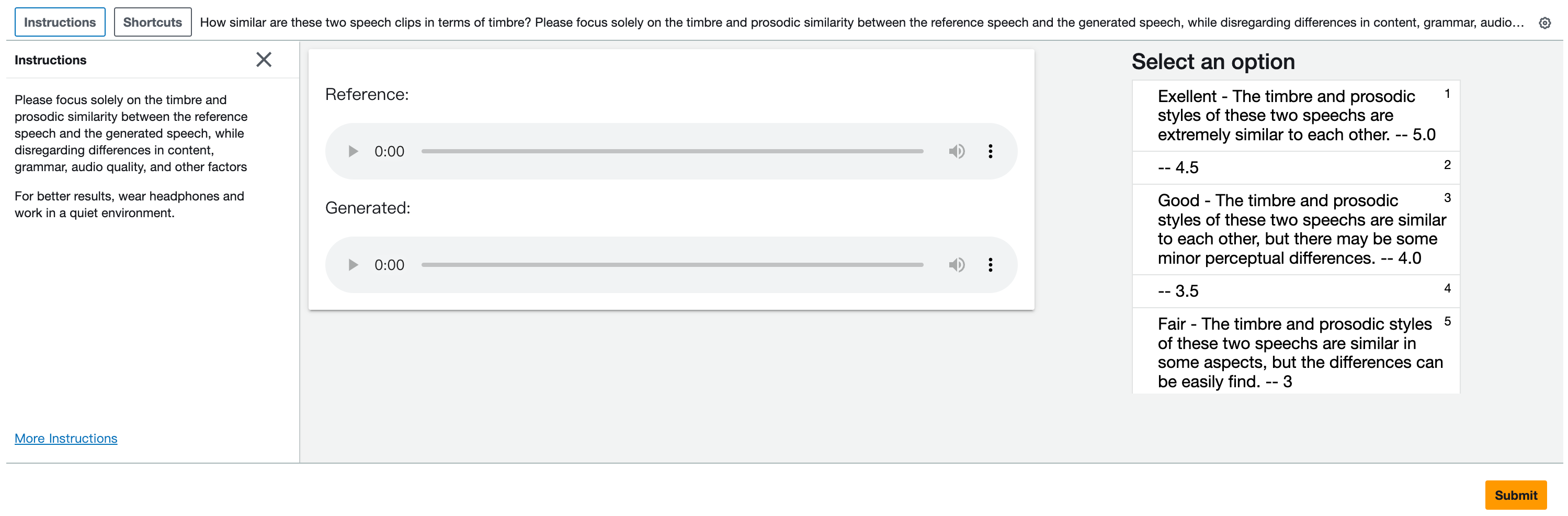}
		\caption*{(b) Screenshot of SMOS testing.}
	\end{minipage}
	\centering
	\begin{minipage}{0.85\linewidth}
		\centering
		\includegraphics[width=1\linewidth]{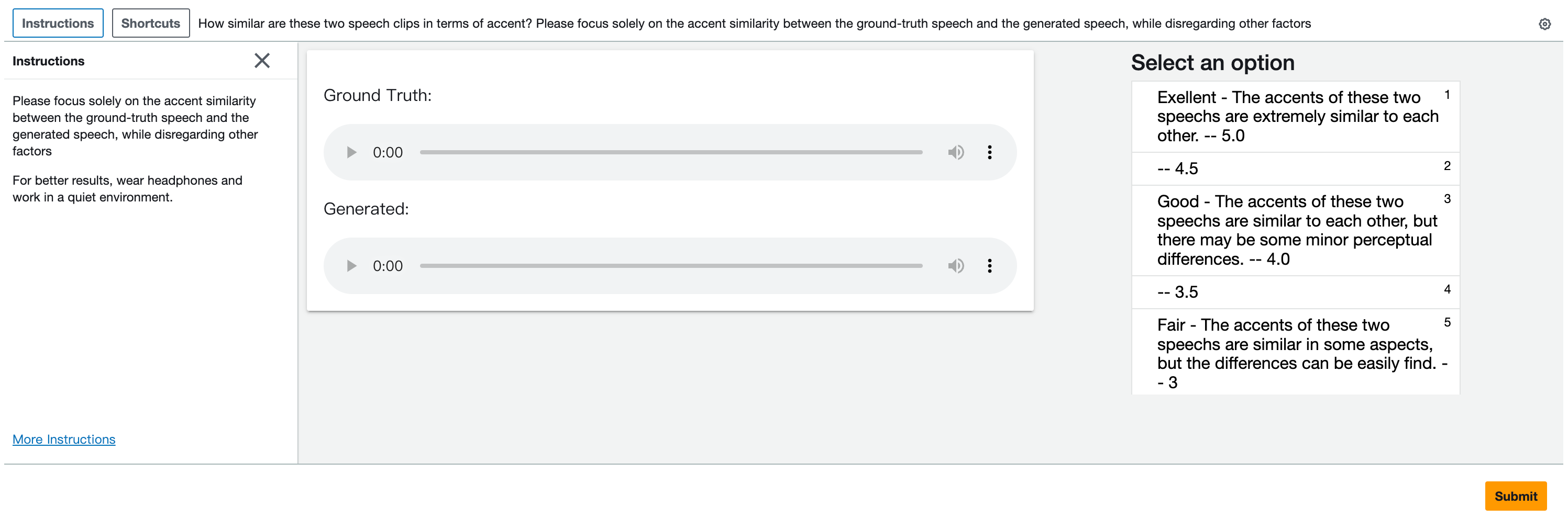}
		\caption*{(c) Screenshot of ASMOS testing.}
	\end{minipage}
	\centering
	\caption{Screenshots of subjective evaluations.}
	\label{screenshots_subjective_evaluations}
\end{figure*}

We conduct evaluations of audio quality, speaker similarity, and accent similarity on Amazon Mechanical Turk (MTurk). We inform the participants that the data will be utilized for scientific research purposes. For each dataset, 40 samples are randomly selected from the test set, and the TTS systems are then used to generate corresponding audio samples. Each audio sample is listened to by a minimum of 10 listeners. For CMOS, following the approach of ~\citet{loizou2011speech}, listeners are asked to compare pairs of audio generated by systems A and B and indicate their preference between the two. They are then asked to choose one of the following scores: 0 indicating no difference, 1 indicating a slight difference, 2 indicating a significant difference and 3 indicating a very large difference. We instruct listeners to ``\textit{Please focus on speech quality, particularly in terms of clarity, naturalness, and high-frequency details, while disregarding other factors}''. For SMOS and ASMOS, each participant is instructed to rate the sentence on a 1-5 Likert scale based on their subjective judgment. For speaker similarity evaluations (SMOS), we instruct listeners to ``\textit{Please focus solely on the timbre and prosodic similarity between the reference speech and the generated speech, while disregarding differences in content, grammar, audio quality, and other factors}''. For accent similarity evaluations (ASMOS), we instruct listeners to ``\textit{Please focus solely on the accent similarity between the ground-truth speech and the generated speech, while disregarding other factors}''. The screenshots of instructions for testers are shown in Figure~\ref{screenshots_subjective_evaluations}. Additionally, we insert audio samples with known quality levels (e.g., reference recordings with no artifacts or intentionally corrupted audio with noticeable distortions) into the evaluation set to verify whether evaluators are attentive and professional. We also randomly repeat some audio clips in the evaluation set to check whether evaluators provide consistent ratings for the same sample. If large deviations in scores (larger than 1.0) for repeated clips occurs, we will select a new rater to evaluate this audio clip. We paid \$8 to participants hourly and totally spent about \$500 on participant compensation.

\section{Classifier-Free Guidance Used in Zero-Shot TTS}
\label{app:CFG_in_zs_tts}
Classifier-Free Guidance (CFG)~\citep{ho2022classifier} is a technique that balances sample fidelity and mode coverage in diffusion models by combining the score estimates from both a conditional and an unconditional model. The unconditional model is trained alongside the conditional model by randomly omitting the conditioning variable $c$ with a certain probability, allowing the same model to provide score estimates for both $p(x)$ and $p(x|c)$. In large-scale zero-shot TTS, VoiceBox~\citep{le2023Voicebox} and NaturalSpeech 2~\citep{shen2023naturalspeech} achieve CFG mechanism by dropping the text and prompt speech features. However, these works overlook that text and timbre should be controlled separately. Inspired by VoiceLDM~\citep{lee2024voiceldm} that introduces separate control of environmental conditions and speech contents, a concurrent work~\citep{yang2024dualspeech} proposes separately controlling the speaker fidelity and text intelligibility. However, this work is limited to improving the audio quality of TTS and does not explore the impact of CFG on accent.

\section{Details of PeRFlow Training Procedure}
\label{app:details_perflow_training}
Once the pretrained ODE solver of the teacher model $\phi_{\theta}$ is available, we perform the PeRFlow technique to train an accelerated solver in real time. When training, we only consider the shortened segments of the ODE trajectories, reducing the computational load of inference for the teacher model at each training step, and accelerating the training process. 

At each training step, given a data sample $z_1$ and a sample $z_0$ drawn from the source distribution (in this case, $z_0 \sim \mathcal{N}(0, I)$, i.e., Gaussian distribution), we randomly select a time window $(t_{k-1}, t_{k}]$ and compute the standpoint of the segmented probability path $z_{t_{k-1}} = \sqrt{1 - \sigma^2(t_{k-1})} z_1 + \sigma(t_{k-1}) z_0$, where $K$ is a hyperparameter indicating the total number of segments, $k\in\{1,\cdots,K\}$, $t_k = k / K$, and $\sigma(t)$ is the noise schedule. The teacher solver only needs to infer the endpoint of this segmented path, $\hat{z}_{t_k} = \phi_{\theta}(z_{t_{k-1}}, t_{k-1}, t_{k})$, with a remarkably smaller number of iterations $\widehat{T}$, comparing to that of a full trajectory, $T$. Finally, the student model is optimized on the segmented trajectory from $z_{t_{k-1}}$ to $\hat{z}_{t_k}$. We set $T$ to 25 and $\widehat{T}$ to 8, achieving a non-negligible acceleration of the training process.

\section{Details about Data and Model Scaling Experiments}
\label{app:data_model_scaling}

\paragraph{Training Corpus.} The data/model scalability is crucial for practical TTS systems. To evaluate the scalability of MegaTTS 3 in Section~\ref{exp:ablation_studies}, we construct a 600kh internal multilingual training corpus comprising both English and Chinese speech. Most of the audiobook recordings are crawled from YouTube and online podcasts like novelfm\footnote{\url{https://novelfm.changdunovel.com/}}. We also include the academic datasets like LibriLight~\citep{kahn2020libri}, WenetSpeech~\citep{zhang2022wenetspeech}, and GigaSpeech~\citep{chen2021gigaspeech}. Since the crawled corpus may contain unlabelled speeches. We transcribe them using an internal ASR model. 

\paragraph{Test Set.} Most prior studies of zero-shot TTS evaluate performances using the reading-style LibriSpeech test set, which may be different from real-world speech generation scenarios. In section~\ref{exp:ablation_studies}, we evaluate our model using the test sets collected from various sources, including: 1) CommonVoice~\citep{ardila2019common}, a large voice corpus containing noisy speeches from various scenarios; 2) RAVDESS~\citep{livingstone2018ryerson}, an emotional TTS dataset featuring 8 emotions and 2 emotional intensity. We follow~\citet{ju2024naturalspeech} and use strong-intensity samples to validate the model’s ability to handle emotional variance; 3) LibriTTS~\citep{zen2019libritts}, a high-quality speech corpus; 4) we collect samples from videos, movies, and animations to test whether our model can simulate timbres with distinctly strong individual characteristics. The test set consists of 40 audio samples extracted from each source.

\paragraph{Experimental Setup} We scale up MegaTTS 3 from 0.5B to 7.0B following the hyper-parameter settings in Qwen 2~\citep{yang2024qwen2}. In this experiment, we only increase the parameters of the MegaTTS 3 model to verify its scalability. The parameters of the speech compression VAE remained unchanged. In theory, expanding the parameters of both models could yield the optimal results, which we leave for future work.

\paragraph{Speech-Text Alignment Labels for Large-Scale Data.} Training an MFA model directly on a 600k-hour dataset is impractical. Therefore, we randomly sampled a 10k-hour subset from the dataset to train a robust MFA model, which is then used to align the full dataset. Since data processing inherently requires some alignment model (such as an ASR model) for speech segmentation, using a pretrained MFA model for alignment extraction does not limit the system's data scalability.

\begin{table}[!t]
\small
\centering
\begin{tabular}{@{}l|cc@{}}
\toprule
\bfseries Setting & \bfseries SIM-O$\uparrow$ & \bfseries WER$\downarrow$ \\       
\midrule
2kh   & 0.52           & 4.27\%                 \\
40kh  & 0.63           & 2.98\%                 \\
200kh & 0.65           & 2.34\%                 \\
600kh & \bfseries 0.66 & \bfseries 2.10\%       \\
\midrule
0.5B & 0.66           & 2.10\%                  \\
1.5B & 0.72           & 1.98\%                  \\
7.0B & \bfseries 0.74 & \bfseries 1.90\%        \\
\bottomrule
\end{tabular}
\caption{Results of data and model scaling experiments.}
\label{table:ablation_scalability}
\end{table}
\paragraph{Results} We evaluate the effectiveness of data and model scaling for the proposed MegaTTS 3 model. In this experiment, we train models with 0.5B parameters on multilingual internal datasets with data sizes of 2kh, 40kh, 200kh, and 600kh, respectively. We also train models with 0.5B, 1.5B, and 7.0B parameters on the 600kh dataset. We evaluate the zero-shot TTS performance in terms of speaker similarity (Sim-O) and speech intelligibility (WER) on an internal test set consisting of 400 speech samples from various sources. Based on Table~\ref{table:ablation_scalability}, we conclude that: 1) as the data size increases from 2kh to 600kh, both the model's speaker similarity and speech intelligibility improve consistently, demonstrating strong data scalability of our model; 2) as the model size scales from 0.5B to 7.0B parameters, SIM-O improves by 12.1\% and WER decreases by 9.52\%, validating the model scalability of MegaTTS 3. Additionally, we find that increasing the model parameters enhances its para-linguistic capabilities, with specific audio examples available on the demo page.

\begin{figure*}[!ht]
\centering
\begin{minipage}{0.49\linewidth}
    \centering
\includegraphics[width=1\linewidth]{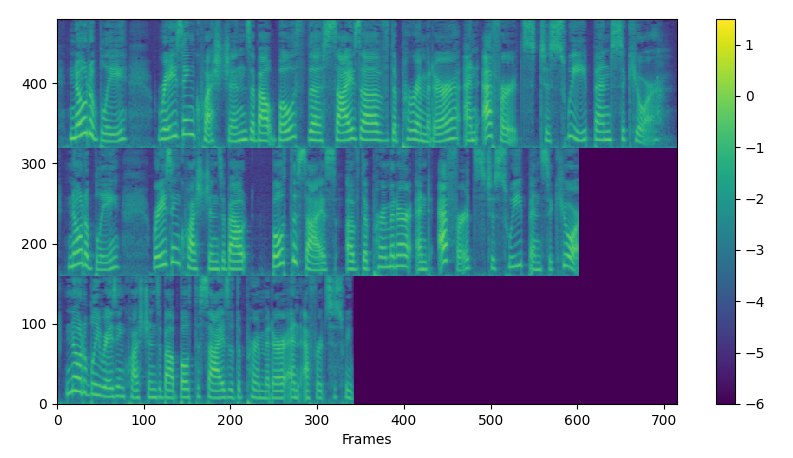}
\caption{Sentence-level duration control.}
\label{app:vis_sent_dur_control}
\end{minipage}
\centering
\begin{minipage}{0.49\linewidth}
    \centering
    \includegraphics[width=1\linewidth]{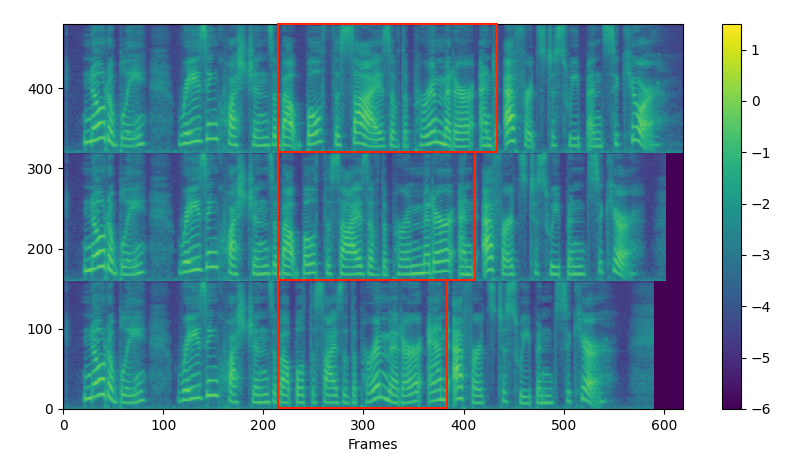}
    \caption{Phoneme-level duration control.}
    \label{app:vis_ph_dur_control}
\end{minipage}
\centering
\end{figure*}

\section{Duration Controllability of MegaTTS 3}
\label{app:dur_contol}
In this section, we aim to verify MegaTTS 3's duration control capabilities through case studies. We randomly selected a speech prompt from the test set and used the sentence ``Notably, raising questions about both the size of the perimeter and efforts to sweep and secure.'' as the target sentence to generate speeches. In the generation process, we first control the sentence-level duration by multiplying the time coordinates of the phoneme anchors described in Section~\ref{method:sec_3_2} by a fixed value. As shown in Figure~\ref{app:vis_sent_dur_control}, our MegaTTS 3 demonstrates good sentence-level duration control. Moreover, our MegaTTS 3 is also capable of fine-grained phoneme-level duration control. As illustrated in Figure~\ref{app:vis_ph_dur_control}, we multiplied the anchor coordinates of the phoneme within the red box by a fixed value while keeping the relative positions of other phoneme anchors unchanged. The figure shows that our MegaTTS 3 also exhibits good fine-grained phoneme-level duration controllability.

\begin{figure*}[!ht]
\centering
\begin{minipage}{0.98\linewidth}
    \centering
\includegraphics[width=1\linewidth]{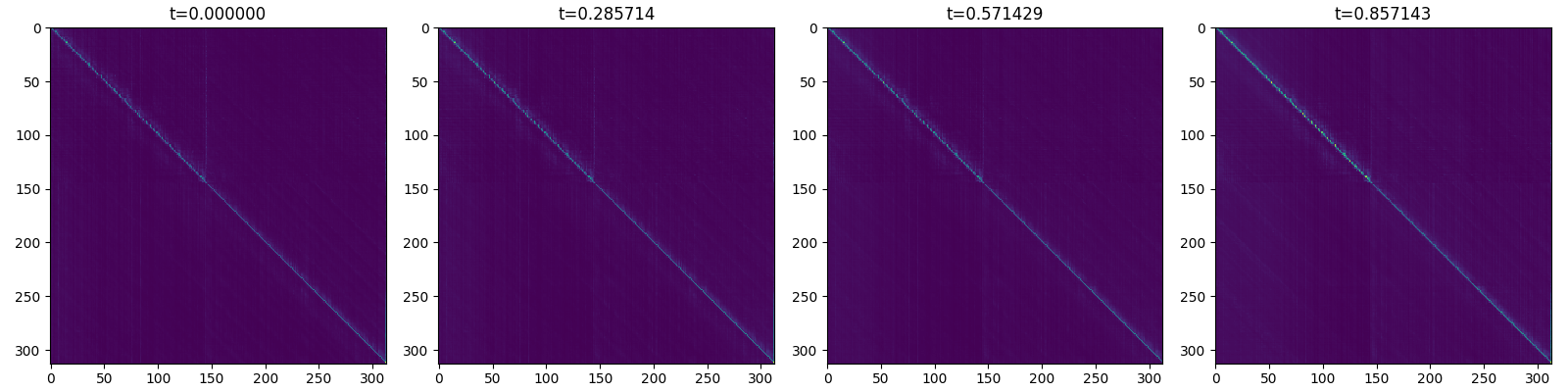}
    \caption*{(a) Layer 8 with different timesteps.}
\end{minipage}
\begin{minipage}{0.98\linewidth}
    \centering
\includegraphics[width=1\linewidth]{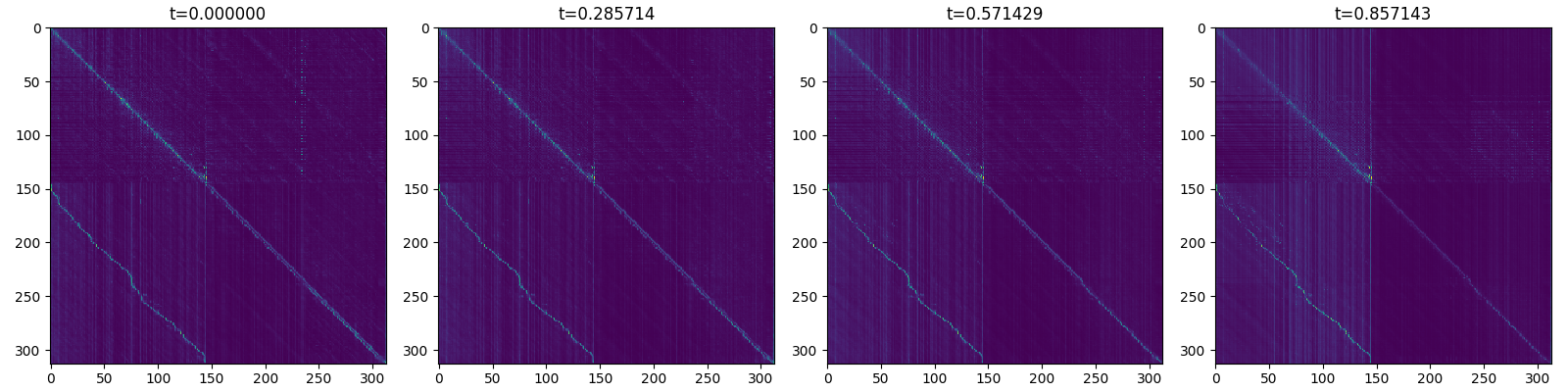}
    \caption*{(b) Layer 16 with different timesteps.}
\end{minipage}
\begin{minipage}{0.98\linewidth}
    \centering
\includegraphics[width=1\linewidth]{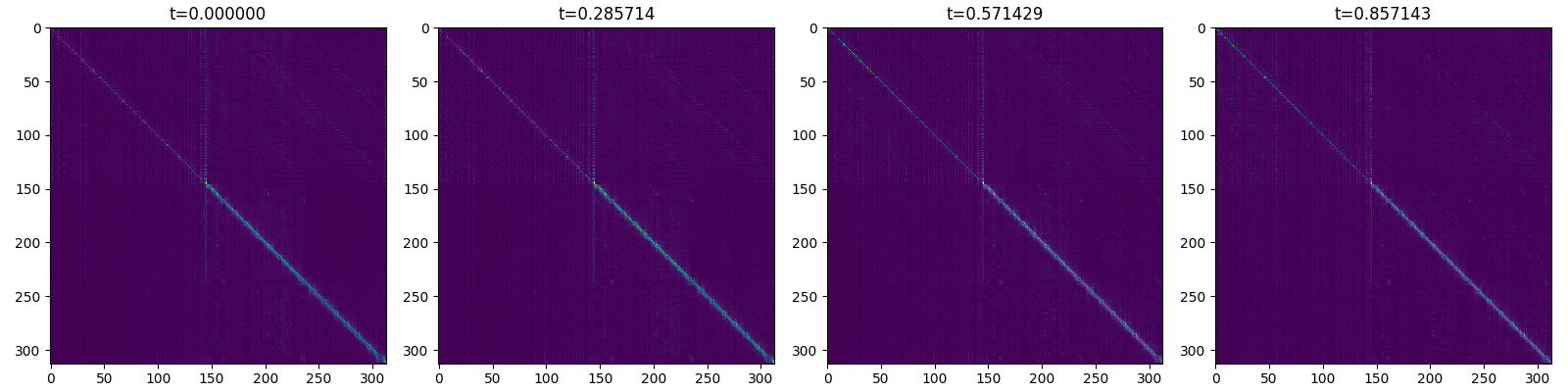}
    \caption*{(c) Layer 27 with different timesteps.}
\end{minipage}
\centering
\caption{Visualization of Attention Matrices from different layers in MegaTTS 3.}
\label{app:vis_attn}
\end{figure*}

\section{Visualization of Attention Matrices}
\label{app:vis_diff_attn}
We visualize the attention matrices from all layers in the 1.4B MegaTTS 3 model, using 8 sampling steps. From Figure~\ref{app:vis_attn}, we observe: 1) within the same layer, despite different timesteps, the attention matrices remain identical. In other words, the function of each layer stays consistent across timesteps; 2) the functions of the transformer layers can be categorized into three types. As shown in Figure~\ref{app:vis_attn} (a), the bottom layers handle text and audio feature extraction; in Figure~\ref{app:vis_attn} (b), the middle layers focus on speech-text alignment; and in Figure~\ref{app:vis_attn} (c), the top layers refine the target latent features.

\begin{table*}[!ht]
\small
\centering
\begin{tabular}{@{}l|cccccc@{}}
\toprule
\bfseries Method & \bfseries MCD$\downarrow$ & \bfseries SSIM$\uparrow$ & \bfseries STOI$\uparrow$ & \bfseries GPE$\downarrow$ & \bfseries VDE$\downarrow$ & \bfseries FFE$\downarrow$ \\       
\midrule
Ours w/ Sparse Alignment & \bfseries 4.56 & \bfseries 0.52  & \bfseries 0.62 & \bfseries 0.34 & \bfseries 0.30 & \bfseries 0.35 \\
Ours w/ Forced Alignment &  4.62   & 0.45 & \bfseries 0.62 & 0.42 & 0.34 & 0.40 \\
Ours w/ Standard CFG & 4.59 & 0.51  & 0.61 & 0.36 & 0.32 & 0.37 \\
Ours w/ Standard AR Duration & 4.58 & 0.50  & \bfseries 0.62 & 0.36 & 0.31 & 0.36 \\
\bottomrule
\end{tabular}
\caption{Comparisons about ``expressiveness'' metrics on the LibriSpeech test-clean set.}
\label{table:expressiveness_exp_libritest_all}
\end{table*}

\begin{table*}[!ht]
\small
\centering
\begin{tabular}{@{}l|cc@{}}
\toprule
\bfseries Model - with Longer Texts & \bfseries WER$\downarrow$ & \bfseries SIM-O$\uparrow$ \\       
\midrule
VoiceCraft      &  12.81\% & 0.62  \\
CosyVoice       &  5.52\%  & 0.68 \\
MegaTTS 3           &  \bfseries 2.39\%  & \bfseries 0.70 \\
\midrule
\midrule
\bfseries Model - with Short Texts & \bfseries WER$\downarrow$ & \bfseries SIM-O$\uparrow$ \\  
\midrule
VoiceCraft       &  4.07\%   & 0.58 \\
CosyVoice      &  2.24\%   & 0.62 \\
MegaTTS 3           &  \bfseries 1.82\%   & \bfseries 0.71  \\
\bottomrule
\end{tabular}
\caption{Comparisons with longer texts.}
\label{table:cmp_longer_samples}
\end{table*}

\section{About Different Lengths of Context}
An imbalanced distribution of prompt and target lengths during training can lead to unstable generation performance during inference. For example, if the majority of the sampled data during training consists of 20-second targets, the generation performance for audio with a 40-second target will be worse than that of 20-second targets in inference. To solve the imbalanced distribution issue, we recommend using the following multi-sentence data sampling strategy: we concatenate all audio recordings of the same speaker in the dataset in time order, and then randomly extract audio segments of length $t\sim U(t_{min}, t_{max})$ from the concatenated audio, where $t_{min}$ is the minimum sampling time and $t_{max}$ is the maximum sampling time. Then, following Section~\ref{method:main_arch}, we randomly divide the sampled sequence into a prompt region and a target region. Although we do not use this strategy in our experiments in order to make a fair comparison with other methods, this strategy is effective in practical scenarios.

\section{Experiments of Prosodic Naturalness for Zero-Shot TTS}
\label{app:exp_expressivenees_zs_tts}

We also conduct the ablation studies using the objective metrics MCD, SSIM, STOI, GPE, VDE, and FFE following InstructTTS~\citep{yang2024instructtts} to evaluate the prosodic naturalness of our proposed method. We conduct experiments on the LibriSpeech test-clean 2.2-hour subset (following the setup in VALL-E 2 and Voicebox). The results are shown in the Table below. We compare MegaTTS 3 with the following baselines: 1) ``Ours w/ Forced Alignment'', we replace the sparse alignment with the forced alignment; 2) ``Ours w/ Standard CFG'', we replace the multi-condition CFG with standard CFG; 3) ``Ours w/ Standard AR Duration'', we replace the duration from F-LM with the duration from standard AR duration predictor following SimpleSpeech 2~\citep{yang2024simplespeech2}. The results in Table~\ref{table:expressiveness_exp_libritest_all} show that sparse alignment brings significant improvements, and both multi-condition CFG and F-LM duration contribute positively to the performance.

\section{Experiments with Longer Samples}
\label{app:exp_longer_samples}
To directly compare MegaTTS 3's robustness to long sequences against other AR models, we have conducted experiemnts for a test set with longer samples. Specifically, we randomly select 10 sentences, each containing more than 50 words. For each speaker in the LibriSpeech test-clean set, we randomly chose a 3-second clip as a prompt, resulting in 400 target samples in total. To make our results more convincing, we include strong-performing TTS models, VoiceCraft~\citep{peng2024voicecraft} and CosyVoice (AR+NAR)~\citep{du2024cosyvoice}, as our baselines. The results for longer samples are presented in Table~\ref{table:cmp_longer_samples}. As shown, compared to the baseline systems, MegaTTS 3 does not exhibit a significant decline in speech intelligibility when generating longer sentences, illustrating the effectiveness of the combination of F-LM and MegaTTS 3.

\begin{table*}[!ht]
\small
\centering
\begin{tabular}{@{}l|cccc@{}}
\toprule
\bfseries Model & \bfseries WER$\downarrow$ & \bfseries Substitution$\downarrow$ & \bfseries Deletion$\downarrow$& \bfseries Insertion$\downarrow$ \\       
\midrule
E2-TTS       &  8.49\%   & 3.65\% & 4.75\% & 0.09\% \\
F5-TTS       &  4.28\%   & \bfseries 1.78\% & 2.28\% & 0.22\% \\
MegaTTS 3        &  \bfseries 3.95\%   & 1.80\% & \bfseries 2.07\% & \bfseries 0.08\% \\
\bottomrule
\end{tabular}
\caption{Comparisons with hard sentences. The results of the baselines are infered from offical
checkpoints.}
\label{table:cmp_hard_sentences}
\end{table*}

\section{Experiments with Hard Sentences}
The transcriptions on the LibriSpeech test-clean set are relatively simple since they come from audiobooks. To further indicate the speech intelligibility of different methods, we evaluate our model on the challenging set containing 100 difficult textual patterns from ELLA-V~\citep{song2024ella}. Since the speech prompts used by ELLA-V are not publicly available, we randomly sample 3-second-long speeches in the LibriSpeech test-clean set as speech prompts. For this evaluation, we used the official checkpoint of F5-TTS~\citep{chen2024f5} and the E2-TTS~\citep{eskimez2024e2} inference API provided on F5-TTS's Hugging Face page. We employ Whisper-large-v3 for WER calculation. Based on the results presented in Table~\ref{table:cmp_hard_sentences}, our model shows stronger robustness against hard transcriptions.

\section{Additional Details for Multi-Condition CFG}
\label{app:additional_detials_for_mt_cfg}
In Section~\ref{method:sec_3_2}, regarding the multi-condition CFG technique, the experimental setup for the preliminary experiment for accent control is: fixing $\alpha_{spk}$ at 2.5 and varying $\alpha_{txt}$ from 1.0 to 6.0. Specifically, as $\alpha_{txt}$ increases from 1.0 to 1.5, the generated speeches contains improper pronunciations and distortions. When $\alpha_{txt}$ ranges from 1.5 to 2.5, the pronunciations align with the speaker's accent. Finally, once $\alpha_{txt}$ exceeds 4.0, the generated speech converges toward the standard pronunciation of the target language. Notably, the optimal values for parameters $\alpha_{txt}$ and $\alpha_{spk}$ may vary across different models. The values reported here are specific to the model used in our experiments.

\section{Ethics Statement}
\label{sec:ethics_statement}
The proposed model, MegaTTS 3, is designed to advance zero-shot TTS technologies, making it easier for users to generate personalized speech. When used responsibly and legally, this technique can enhance applications such as movies, games, podcasts, and various other services, contributing to increasing convenience in everyday life. However, we acknowledge the potential risks of misuse, such as voice cloning for malicious purposes. To mitigate this risk, solutions like building a corresponding deepfake detection model will be considered. Additionally, we plan to incorporate watermarks and verification methods for synthetic audio to ensure ethical use in real-world applications. Restrictions will also be included in the licensing of our project to further prevent misuse. By addressing these ethical concerns, we aim to contribute to the development of responsible and beneficial AI technologies, while remaining conscious of the potential risks and societal impact.

\section{Reproducibility Statement}
\label{sec:reproducibility_statement}
We have taken several steps to ensure the reproducibility of the experiments and results presented in this paper: 1) the architecture and algorithm of the MegaTTS 3 model are described in Section~\ref{method} and and relevant hyperparameters are fully described in Appendix~\ref{app:model_config}; 2) The evaluation metrics, including WER, SIM-O, MCD (dB), the moments of the pitch distribution, alignment error, CMOS, SMOS, and ASMOS, are described in detail in Section~\ref{Experimental_Setup}; 3) For most of the key experiments, we utilize publicly available datasets such as LibriLight, LibriSpeech, and L2Arctic. The selection of the test sets is identical to that used in previous zero-shot TTS research. However, as the publicly available datasets are insufficient for our data scaling experiments, we construct a larger dataset, which is described in detail in Appendix~\ref{app:data_model_scaling}; 4) To ensure reproducibility of the results, we have carefully set random seeds in our experiments and the random seeds are provided in Appendix~\ref{app:random_seeds}. All objective results reported are based on the average performance across multiple runs.


            

\end{document}